\documentclass{moriond}

\usepackage{amsmath}
\usepackage{amssymb}

\def\be{\begin{equation}}
\def\ee{\end{equation}}
\def\bea{\begin{eqnarray}}
\def\eea{\end{eqnarray}}

\newcommand{\Photo}{}

\begin{document}
\title{Scalar Induced Gravitational Waves signaling Primordial Black Hole Dark Matter}

\author{Cristian Joana}


\address{
Institute of Theoretical Physics,
Chinese Academy of Sciences, Beijing 100190, China
}

\maketitle
\abstracts{
Primordial black holes are a unique probe of the early Universe and offer a potential link between inflationary dynamics and dark matter. In a recent work~\cite{Inui:2024fgk}, we analyse PBH formation and the associated stochastic gravitational wave background from scalar induced gravitational waves, while taking into account the contributions from local non-Gaussianities. We highlight the observable consequences for the LISA and PTA experiments.\\
\textbf{Contribution to the 2025 Gravitation session of the 59th Rencontres de Moriond.}
}

\setlength{\parindent}{0pt}

\section{Introduction}
Primordial black holes (PBHs) are compelling probes of the primordial Universe, capable of constraining inflationary models through their sensitivity to rare, large-amplitude curvature perturbations. In the standard picture, these rare overdensities might collapse to form a PBH during the early radiation-dominated era.
These same fluctuations inevitably source a stochastic background of scalar-induced gravitational waves (SIGWs), which would lie within the sensitivity range of experiments such as LISA/Taiji/TianQin and pulsar timing array (PTA) observations.

\section{Computation of PBH abundances and their SIGWs}
In our recent work~\cite{Inui:2024fgk}, we explore PBH formation and SIGW production in a class of inflationary scenarios with intrinsic local non-Gaussianity (NG). In our analysis, we employ the \textit{logarithmic duality}~\cite{Pi:2022ysn}, which links the non-Gaussian parameter $\gamma$ to both the curvature perturbation profile $\zeta(r)$ and its probability distribution function (PDF). This relation allows us to capture the two key effects of NGs on PBH formation:
\begin{enumerate}
    \item The shape of the PDF tail is modified, dramatically impacting the abundance of PBHs due to its exponential sensitivity. The PDF for the NG curvature perturbation reads 
    \begin{equation}
    P(\zeta) = \frac{1}{\sqrt{2 \pi \sigma^2}} \exp\left[ -\frac{1}{2 \gamma^2 \sigma^2} \left( e^{-\gamma \zeta} - 1 \right)^2 - \gamma \zeta \right],       
    \end{equation}
    \item The curvature profile $\zeta(r)$ is altered, affecting the threshold amplitude $\mu_c(\gamma)$ necessary for gravitational collapse. The averaged curvature profile takes the form 
    \begin{equation}
        \zeta(r; \mu, \gamma) = -\frac{1}{\gamma} \ln \left[ 1 - \gamma \zeta_G(r; \mu) \right], \quad \zeta_G(r; \mu) = \mu\, \mathrm{sinc}(k r) ~,
    \end{equation}
    after a local transformation from the Gaussian fluctuation $\zeta_G$ with amplitude $\mu$.
\end{enumerate}

Positive NG ($\gamma > 0$) enhances the PDF tail and sharpens the peak of $\zeta(r)$ at the center, reducing the collapse threshold $\mu_c$ and increasing PBH production. In contrast, negative NG ($\gamma < 0$) suppresses the tail and leads to broader, plateau-like profiles. These profiles shift the Laplacian away from the center, introducing complexity in the Ricci curvature structure—a feature to be explored in future work.
Both of these effects tend to suppress the PBH formation when $0 > \gamma > -3.1$. For strongly negative $\gamma \lesssim -3.1$ values, the collapse threshold $\mu_c$ begins to decrease again, as the off-center Laplacian grows to become energetically dominant. This enhances again the PBH production for these strongly negative $\gamma$-values (see Fig.~1 in Ref.~[1]).  

Our methodology for computing the PBH abundances and the SIGWs is as follows: 
\begin{itemize}
    \item The collapse thresholds $\mu_c(\gamma)$ are computed using full numerical relativity simulations, for curvature perturbations re-entering during the radiation-dominated epoch. 
    \item The PBH abundances are evaluated assuming a monochromatic curvature power spectrum centered at $k_\star$ with amplitude $A_G$; that is ${\cal P}_\zeta(k) = A_G \delta (k_\star - k)~.$ 
    \item The amplitude $A_G$ is fixed by that all dark matter in the form of PBHs (i.e. $f_{\rm PBH}^{\rm tot} = 1$).
    \item Once ${\cal P}_\zeta$ is determined, we compute the SIGWs perturbatively accounting for {NG corrections}~\cite{Abe:2022xur} up to third order, ${\cal O}(A^3_S)$.
\end{itemize}


\begin{figure*}[h!]
\begin{minipage}{0.99\linewidth}
\centerline{
\includegraphics[width=0.9\linewidth]{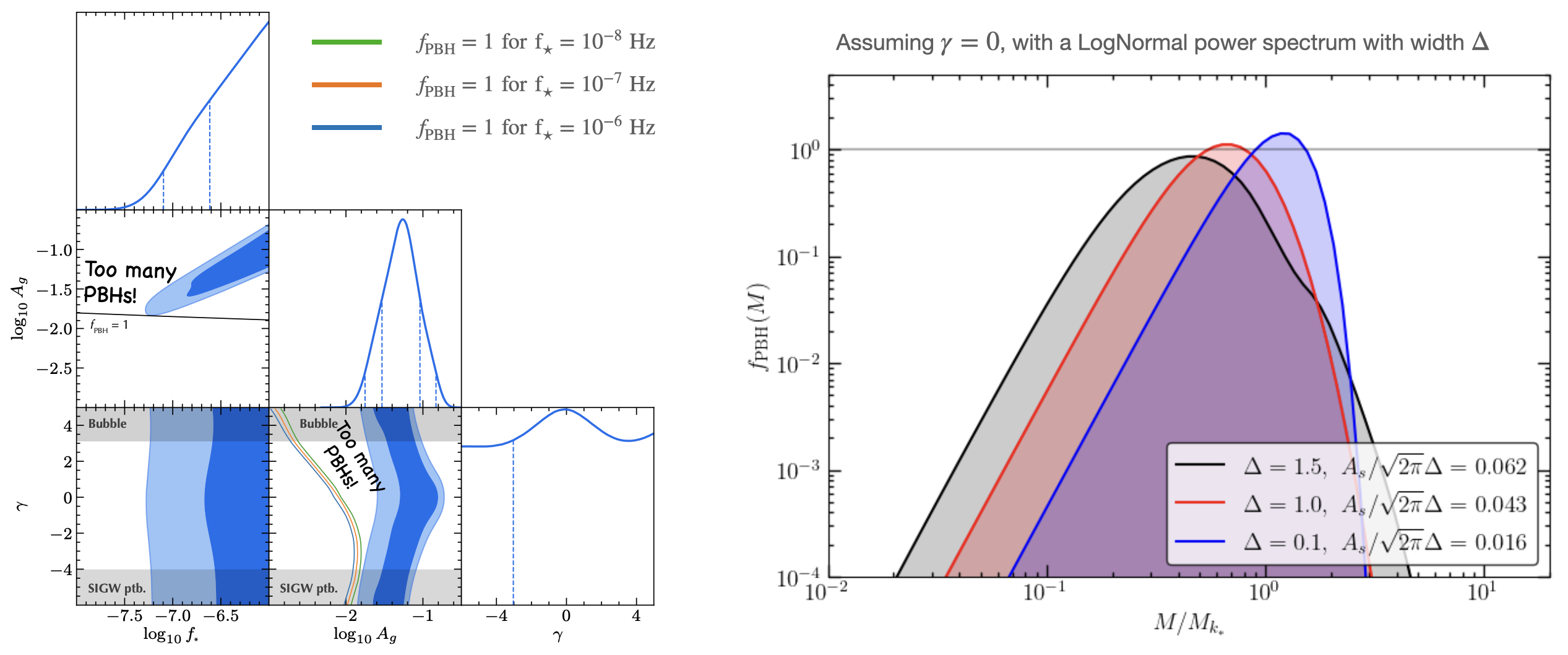}
}
\end{minipage}
\caption[]{
\textit{Left:} The posterior plot generated by the NANOGrav result where SIGWs include
contributions up to ${\cal O}(A^3_S)$. This figure is adapted from Ref.~[1]. 
\textit{Right:} Preliminary PBH mass function computed using the peak theory methodology~\cite{Pi:2024ert}, assuming $f^{\rm tot}_{\rm PBH} = 1$ and Gaussian statistics ($\gamma \approx f_{\rm NL} \approx 0$). For widths larger than $\Delta > 0.1$ we find peak amplitudes above $A_s/\sqrt{2\pi}\Delta \gtrsim 3 \cdot 10^{-2}$ which can explain the reported PTA signal~\cite{NANOGrav:2023hvm}. 
}
\label{fig:PBHAbundances}
\end{figure*}

\section{Connection with LISA and PTA experiments}

We then evaluate the SIGW spectrum in both the mHz band (relevant for LISA) and the nHz band (relevant for PTAs); see Ref.~[1] for full results and a detailed discussion. Notably, we find that the PTA signal~\cite{NANOGrav:2023hvm} is incompatible with PBH formation scenarios under the assumption of a monochromatic power spectrum, as the detected amplitude would imply an overproduction of PBHs---exceeding the expected dark matter abundance. However, preliminary analyses indicate that this tension can be alleviated when one considers a finite-width peak in the power spectrum (e.g., a lognormal peak with width $\Delta$). Fig.~1 left-panel illustrates the posterior estimation from the NANOGrav results~\cite{NANOGrav:2023hvm}, ruling out a monochromatic power spectrum.
The right panel shows that larger power spectrum values become accessible when widening the peak width, tentatively  resolving the overproduction problem even for nearly Gaussian perturbations (i.e. $\gamma \approx 0$).

\section*{Acknowledgments}
C.J. is supported by NSFC grants No. W2433007 and 12475066. C.J. acknowledges contributions from Jianing Wang,  Shi Pi, Ryoto Inui, Yuichiro Tada, Shuichiro Yokoyama, and Hayato Motohashi. 
C.J. thanks Mauro Pieroni and Alberto Ropert Pol for stimulating discussions during the conference, as well as to the Moriond 2025 organisers for their hospitality.

\end{document}